\documentclass[12pt]{article}
\usepackage{amsmath,amsfonts,amssymb}

\begin{document}
\thispagestyle{empty}

\def\theequation{\arabic{section}.\arabic{equation}}
\def\a{\alpha}
\def\b{\beta}
\def\g{\gamma}
\def\d{\delta}
\def\dd{\rm d}
\def\e{\epsilon}
\def\ve{\varepsilon}
\def\z{\zeta}
\def\B{\mbox{\bf B}}

\newcommand{\h}{\hspace{0.5cm}}

\begin{titlepage}
\vspace*{1.cm}
\renewcommand{\thefootnote}{\fnsymbol{footnote}}
\begin{center}
{\Large \bf Leading finite-size effects on some three-point
correlators in $TsT$-deformed $AdS_5\times S^5$}
\end{center}
\vskip 1.2cm \centerline{\bf Plamen Bozhilov} \vskip 0.6cm
\centerline{\sl Institute for Nuclear Research and Nuclear Energy}
\centerline{\sl Bulgarian Academy of Sciences} \centerline{\sl
1784 Sofia, Bulgaria}

\centerline{\tt plbozhilov@gmail.com}

\vskip 20mm

\baselineskip 18pt

\begin{center}
{\bf Abstract}
\end{center}
\h We compute the leading finite-size effects on the normalized
structure constants in semiclassical three-point correlation
functions of two finite-size giant magnon string states and three
different types of "light" states - primary scalar operators,
dilaton operator with nonzero momentum and singlet scalar
operators on higher string levels. This is done for the case of
$TsT$-transformed, or $\gamma$-deformed, $AdS_5\times S^5$ string
theory background, dual to $\mathcal{N} = 1$ super Yang-Mills
theory in four dimensions, arising as an exactly marginal
deformation of $\mathcal{N} = 4$ super Yang-Mills.

\end{titlepage}
%\end{quote}
%\vskip 1cm \centerline{\today}
\newpage
%\baselineskip 18pt

%%%%%%%%%%%%%%%%%%%%%%%%%%%%%%%%%%%%%%%%%%%%%%%%%
\def\nn{\nonumber}
%%%%%%%%%%%%%%%%%%%%%%%%%%%%%%%%%%%%%%%%%%%%%%%%%
%%%%%%%%%%%%%%%%%%%%%%%%%%%%%%%%%%%%%%%%%%%%%%%%%%%%%
\def\tr{{\rm tr}\,}
\def\p{\partial}
\newcommand{\bea}{\begin{eqnarray}}
\newcommand{\eea}{\end{eqnarray}}
\newcommand{\bde}{{\bf e}}
\renewcommand{\thefootnote}{\fnsymbol{footnote}}
\newcommand{\be}{\begin{equation}}
\newcommand{\ee}{\end{equation}}
%\newcommand{\h}{\hspace{0.5cm}}
%%%%%%%%%%%%%%%%%%%%%%%%%%%%%%%%%%%%%%%%%%%%%%%%%%%%

\vskip 0cm

\renewcommand{\thefootnote}{\arabic{footnote}}
\setcounter{footnote}{0}

%\setcounter{equation}{0}
%\section{Introduction}

\setcounter{equation}{0}
%%%%%%%%%%%%%5%%%%%%%%%%%%%%%%%%%%%%%%%%%%%%%%%%%%%%%%%%%%%%%%%%%%%%%%%%%%%%%%
\section{Introduction}
%%%%%%%%%%%%%5%%%%%%%%%%%%%%%%%%%%%%%%%%%%%%%%%%%%%%%%%%%%%%%%%%%%%%%%%%%%%%%%
The AdS/CFT duality \cite{AdS/CFT} between string theories on
curved space-times with Anti-de Sitter subspaces and conformal
field theories in different dimensions has been actively
investigated in the last years. A lot of impressive progresses
have been made in this field of research based mainly on the
integrability structures discovered on both sides of the
correspondence (for recent overview on the AdS/CFT duality, see
\cite{RO}). The most studied example is the correspondence between
type IIB string theory on ${\rm AdS}_5\times S^5$ target space and
the ${\cal N}=4$ super Yang-Mills theory (SYM) in four space-time
dimensions. However, many other cases are also of interest, and
have been investigated intensively.

Different classical string solutions play important role in
checking and understanding the AdS/CFT correspondence
\cite{AAT1012}. To establish relations with the dual gauge theory,
one has to take the semiclassical limit of {\it large} conserved
charges \cite{GKP02}. A crucial example of such string solution is
the so called "giant magnon", discovered by Hofman and Maldacena
in $R_t\times S^2$ subspace of ${\rm AdS}_5\times S^5$
\cite{HM06}. It gave a strong support for the conjectured all-loop
$SU(2)$ spin chain, arising in the dual $\mathcal{N}=4$ SYM, and
made it possible to get a deeper insight in the AdS/CFT duality.
Characteristic feature of this solution is that the string energy
$E$ and the angular momentum $J_1$ go to infinity, but the
difference $E-J_1$ remains finite and related to the momentum of
the magnon excitations in the dual spin chain in $\mathcal{N}=4$
SYM. This string configuration have been extended to the case of
{\it dyonic} giant magnon, being solution for a string moving on
$R_t\times S^3$ and having second nonzero angular momentum $J_2$
\cite{ND06}. Further extension to $R_t\times S^5$ have been also
worked out in \cite{KRT06}. It was also shown there that such type
of string solutions can be obtained by reduction of the string
dynamics to the Neumann-Rosochatius integrable system, by using a
specific ansatz.

An interesting issue to solve is to find the {\it finite-size
effect}, i.e. $J_1$ large, but finite, related to the wrapping
interactions in the dual field theory \cite{Janikii}. For (dyonic)
giant magnons living in ${\rm AdS}_5\times S^5$ this was done in
\cite{AFZ06,HS0801}. The corresponding string solutions, along
with the (leading) finite-size corrections to their dispersion
relations have been found.

Here, we are going to consider the leading finite-size effects on
some three-point correlation functions in $\gamma$-deformed
\cite{LM05} or $TsT$-transformed \cite{F05} $AdS_5\times S^5$
string theory background. To this end, we will need to use our
knowledge about the properties of the finite-size (dyonic) giant
magnon solutions on this target space. The corresponding
information can be found in \cite{BF08,AB2010}.

In this paper we will be interested in the case of three-point
correlators, when two of the "heavy" string states are finite-size
giant magnons, while the third state is a "light" one\footnote{The
first papers in which three-point correlation functions of two
"heavy" operators and a "light" operator have been computed are
\cite{Zarembo, Costa}.}. We will consider three different types of
"light" states: primary scalar operators, dilaton operator with
nonzero momentum and singlet scalar operators on higher string
levels. The finite-size effects on such correlation functions in
$TsT$-transformed $AdS_5\times S^5$ have been found in
\cite{PLB1107,PLB1108}. There, the normalized structure constants
in these correlators are given in terms of several parameters and
hypergeometric functions of two variables depending on them. On
the other hand, it is important to know their dependence on the
conserved string charges $J_1$, $J_2$ and the worldsheet momentum
$p$, because namely these quantities are related to the
corresponding operators in the dual $\mathcal{N} = 1$ SYM, and the
momentum of the magnon excitations in the dual spin-chain. That is
why, we are going to find this dependence here. Unfortunately,
this can not be done exactly for the finite-size case due to the
complicated dependence between the above mentioned parameters and
$J_1$, $J_2$, $p$. Because of that, we will consider only the
leading order finite-size effects on the three-point correlators
of this type. Moreover, due to computational complications, we
will restrict ourselves to the case of $J_2=0$.

This paper is organized as follows. In Sec. 2, we give a short
review of the finite-size (dyonic) giant magnon's solution on
$\gamma$-deformed $AdS_5\times S^5$. Also, we give the
corresponding exact semiclassical results for the three-point
correlators we are interested in, found in \cite{PLB1107,PLB1108}.
Sec. 3 is devoted to the computation of the leading order
finite-size effects on the three-point correlators given in Sec. 2
in terms of the conserved string angular momentum $J_1$ and the
worldsheet momentum $p$. In Sec. 4 we conclude with some final
remarks.

\setcounter{equation}{0}
\section{Finite-size giant magnons\\ on $TsT$-deformed $AdS_5\times S^5$ and
\\some three-point correlators}

\subsection{Short review of the giant magnon solutions}

Investigations on AdS/CFT duality for the cases with reduced or
without supersymmetry is of obvious importance and interest. An
example of such correspondence between gauge and string theory
models with reduced supersymmetry is provided by an exactly
marginal deformation of $\mathcal{N} = 4$ SYM and string theory on
a $\beta$-deformed $AdS_5\times S^5$ background suggested by Lunin
and Maldacena in \cite{LM05}. When $\beta\equiv\gamma$ is real,
the deformed background can be obtained from $AdS_5\times S^5$ by
the so-called TsT transformation on $S^5$. It includes T-duality
on one angle variable, a shift of another isometry variable, then
a second T-duality on the first angle \cite{F05}. The $AdS_5$ part
of the background is untouched, so the conformal invariance
remains.

An essential property of the TsT transformation is that it
preserves the classical integrability of string theory on
$AdS_5\times S^5$ \cite{F05}. The $\gamma$-dependence enters only
through the {\it twisted} boundary conditions and the {\it
level-matching} condition. The last one is modified since a closed
string in the deformed background corresponds to an open string on
$AdS_5\times S^5$ in general.

The parameter $\tilde{\gamma}$, which appears in the string
action, is related to the deformation parameter $\gamma$ as
\bea\nn \tilde{\gamma}= 2\pi T \gamma= \sqrt{\lambda}\ \gamma
,\eea where $T$ is the string tension and $\lambda$ is the t'Hooft
coupling.

The effect of introducing $\gamma$ on the field theory side of the
duality is to modify the super potential as follows \bea\nn
W\propto
tr\left(e^{i\pi\gamma}\Phi_1\Phi_2\Phi_3-e^{-i\pi\gamma}\Phi_1\Phi_3\Phi_2\right).\eea
This leads to reduction of the supersymmetry of the SYM theory
from $\mathcal{N}=4$ to $\mathcal{N}=1$.

Since we are going to consider three-point correlation functions
with two vertices corresponding to giant magnon states, we can
restrict ourselves to the subspace $R_t\times S_{\gamma}^3$ of
$AdS_5\times S_{\gamma}^5$ background. Then one can show that by
using the ansatz \bea\label{NRA} &&t(\tau,\sigma)=\kappa\tau,\h
\theta(\tau,\sigma)=\theta(\xi),\h
\phi_j(\tau,\sigma)=\omega_j\tau+f_j(\xi),\\ \nn
&&\xi=\alpha\sigma+\beta\tau,\h \kappa, \omega_j, \alpha,
\beta=constants,\h j=1,2,\eea the string Lagrangian in conformal
gauge, on the $\gamma$-deformed three-sphere $S_{\gamma}^3$, can
be written as \cite{AB11062} (prime is used for $d/d \xi$) \bea\nn
&&\mathcal{L}_\gamma=(\alpha^2-\beta^2)
\left[\theta'^2+G\sin^2\theta\left(f'_1-\frac{\beta\omega_1}{\alpha^2-\beta^2}\right)^2
+G\cos^2\theta\left(f'_2-\frac{\beta\omega_2}{\alpha^2-\beta^2}\right)^2
\right.
\\ \label{r2} &&-\left.\frac{\alpha^2}{(\alpha^2-\beta^2)^2}G
\left(\omega_1^2\sin^2\theta+\omega_2^2\cos^2\theta\right)
+2\alpha\tilde{\gamma}G\sin^2 \theta \cos^2 \theta \frac{\omega_2
f'_1-\omega_1 f'_2}{\alpha^2-\beta^2}\right], \eea where \bea\nn
G=\frac{1}{1+\tilde{\gamma}^2\sin^2 \theta \cos^2 \theta}.\eea

By using (\ref{r2}) and the Virasoro constraints, one can find the
following first integrals of the string equations of motion
\bea\nn &&f'_1=\frac{\Omega_1}{\alpha}\frac{1}{1-v^2}
\left[\frac{v W-u K}{1-\chi}
-v(1-\tilde{\gamma}K)-\tilde{\gamma}u\chi\right],
\\ \label{tfid} &&f'_2=\frac{\Omega_1}{\alpha}\frac{1}{1-v^2} \left[\frac{K}{\chi}
-uv(1-\tilde{\gamma}K)-\tilde{\gamma}v^2W+\tilde{\gamma}(1-\chi)\right],
\\ \nn && \chi'=
\frac{2\sqrt{1-u^{2}}}{1-v^2}
\sqrt{(\chi_{p}-\chi)(\chi-\chi_{m})(\chi-\chi_n)},\eea where
\bea\nn &&\chi=\cos^2\theta,\h v=-\beta/\alpha,\h
u=\frac{\Omega_2}{\Omega_1},\h
W=\left(\frac{\kappa}{\Omega_1}\right)^2,\h
K=\frac{C_2}{\alpha\Omega_1},
\\ \nn &&\Omega_1=\omega_1\left(1+\tilde{\gamma}\frac{C_2}{\alpha\omega_1}\right), \h
\Omega_2=\omega_2\left(1-\tilde{\gamma}\frac{C_1}{\alpha\omega_2}\right)
,\h C_1,\ C_2=constants.\eea Also, the following equalities hold
\bea\nn &&\chi_p+\chi_m+\chi_n=\frac{2-(1+v^2)W-u^2}{1
-u^2},\\
\label{3eqsd} &&\chi_p \chi_m+\chi_p \chi_n+\chi_m
\chi_n=\frac{1-(1+v^2)W+(v W-u K)^2-K^2}{1 -u^2},\\ \nn && \chi_p
\chi_m \chi_n=- \frac{K^2}{1 -u^2}.\eea
The case of dyonic
finite-size giant magnons corresponds to \bea\nn 0<u<1,\h 0<v<1,\h
0<W<1,\h 0<\chi_{m}<\chi< \chi_{p}<1,\h \chi_{n}<0.\eea

The $AdS_5$ part of the giant magnon solution, in Euclidean
Poincare coordinates, can be written as\footnote{Euclidean
continuation of the time-like directions to $t_e = it$, $x_{0e} =
ix_0$, will allow the classical trajectories to approach the
$AdS_5$ boundary $z=0$ when $\tau_e \rightarrow \pm\infty$, and to
compute the corresponding correlation functions.}
($t=\sqrt{W}\tau$,\ $i\tau=\tau_e$)\footnote{We set
$\alpha=\Omega_1=1$ for simplicity.} \bea\nn
&&z=\frac{1}{\cosh(\sqrt{W}\tau_e)}\h
x_{0e}=\tanh(\sqrt{W}\tau_e),\h x_i=0,\h i=1,2,3.\eea

Let us also write down the exact expressions for the conserved
charges and the angular differences \bea\label{E}
&&\mathcal{E}\equiv \frac{2\pi E}{\sqrt{\lambda}}
=2\frac{(1-v^2)\sqrt{W}}{\sqrt{1-u^2}}\frac{
\mathbf{K}(1-\epsilon)}{\sqrt{\chi_{p}-\chi_{n}}},
\\ \label{J1} &&\mathcal{J}_1\equiv\frac{2\pi J_1}{\sqrt{\lambda}}=\frac{2}{\sqrt{1-u^2}}
\left[\frac{1-\chi_n-v\left(v W-u
K\right)}{\sqrt{\chi_{p}-\chi_{n}}}\mathbf{K}(1-\epsilon)-\sqrt{\chi_{p}-\chi_{n}}
\mathbf{E}(1-\epsilon)\right],\\ \label{J2}
&&\mathcal{J}_2\equiv\frac{2\pi
J_2}{\sqrt{\lambda}}=\frac{2}{\sqrt{1-u^2}} \left[\frac{u\chi_n-v
K}{\sqrt{\chi_{p}-\chi_{n}}} \mathbf{K}(1-\epsilon)
+u\sqrt{\chi_{p}-\chi_{n}} \mathbf{E}(1-\epsilon)\right],\eea
\bea \label{p1}
&&p_1\equiv\Delta\phi_1=\phi_1(L)-\phi_1(-L)=\frac{2}{\sqrt{1-u^2}}
\\ \nn &&\times\Bigg\{\frac{v W-u
K}{(1-\chi_p)\sqrt{\chi_{p}-\chi_{n}}}
\Pi\left(-\frac{\chi_{p}-\chi_{m}}{1-\chi_{p}}\vert
1-\epsilon\right)-\left[v\left(1-\tilde{\gamma}K\right)+\tilde{\gamma}u\chi_{n}\right]
\frac{\mathbf{K}(1-\epsilon)}{\sqrt{\chi_{p}-\chi_{n}}}  \\
\nn &&-\tilde{\gamma}u\sqrt{\chi_{p}-\chi_{n}}
\mathbf{E}(1-\epsilon)\Bigg\},\eea \bea
\label{p2}&&p_2\equiv\Delta\phi_2=\phi_2(L)-\phi_2(-L)=\frac{2}{\sqrt{1-u^2}}
\\ \nn &&\times\Bigg\{\frac{K}{\chi_p\sqrt{\chi_{p}-\chi_{n}}}
\Pi\left(1-\frac{\chi_{m}}{\chi_{p}}\vert 1-\epsilon\right) -
\left[uv+\tilde{\gamma}v\left(v W-u
K\right)-\tilde{\gamma}\left(1-\chi_n\right)\right]
\frac{\mathbf{K}(1-\epsilon)}{\sqrt{\chi_{p}-\chi_{n}}}
\\ \nn &&-\tilde{\gamma}
\sqrt{\chi_{p}-\chi_{n}} \mathbf{E}(1-\epsilon)\Bigg\},\eea where
the following notation has been introduced \bea\label{de}
\epsilon=\frac{\chi_{m}-\chi_{n}}{\chi_{p}-\chi_{n}}.\eea Here,
$E$, $J_{1,2}$ are the string energy and angular momenta.
$\mathbf{K}(1-\epsilon)$, $\mathbf{E}(1-\epsilon)$ and

$\Pi\left(1-\frac{\chi_{m}}{\chi_{p}}\vert 1-\epsilon\right)$ are
the complete elliptic integrals of first, second and third kind.
The parameter $L$ appeared above is related to the size of the
giant magnons. For finite-size giant magnons $L$ is finite, while
for infinite-size giant magnons $L\to\infty$.

Let us also point out that for the $\gamma$-deformed case even the
giant magnon with $J_2=0$ lives on $S^3_\gamma$. It happens
because that is the smallest consistent reduction due to the {\it
twisted} boundary conditions \cite{BF08}.

The dyonic giant magnon dispersion relation, including the leading
finite-size correction, can be written as \cite{AB2010}
\bea\label{dfsdr} \mathcal{E}-\mathcal{J}_{1}=
\sqrt{\mathcal{J}_2^2+4\sin^2(p/2)} - \frac{\sin^4(p/2)}
{\sqrt{\mathcal{J}_2^2+4\sin^2(p/2)}}\cos(\Phi)\ \epsilon,\eea
where \bea\label{epsilon} &&\epsilon=16
\exp\left[-\frac{2\left(\mathcal{J}_1 +
\sqrt{\mathcal{J}_2^2+4\sin^2(p/2)}\right)
\sqrt{\mathcal{J}_2^2+4\sin^2(p/2)}\sin^2(p/2)}{\mathcal{J}_2^2+4\sin^4(p/2)}
\right].\eea The second term in (\ref{dfsdr}) represents the
leading finite-size effect on the energy-charge relation, which
disappears for $\epsilon\to 0$, or equivalently
$\mathcal{J}_1\to\infty$. It is nonzero only when $\mathcal{J}_1$
is finite. The $\gamma$-deformation effect is represented by
$\cos(\Phi)$.

In the next section, we will restrict our considerations to the
case $\mathcal{J}_2=0$. Then (\ref{dfsdr}) simplifies to
\bea\label{0} \mathcal{E}-\mathcal{J}_{1}= 2\sin(p/2) -
\frac{1}{2}\sin^3(p/2)\cos(\Phi)\ \epsilon,\eea where
\bea\label{eps0} &&\epsilon=16
\exp\left[-\frac{\mathcal{J}_1}{\sin(p/2)} - 2\right],\h
\Phi=2\pi\left(n_2-\frac{\tilde{\gamma}}{2\pi}\mathcal{J}_1\right),\h
n_2\in \mathbb{Z}.\eea

\subsection{Semiclassical three-point correlation functions}
It is known that the correlation functions of any conformal field
theory can be determined  in principle in terms of the basic
conformal data $\{\Delta_i,C_{ijk}\}$, where $\Delta_i$ are the
conformal dimensions defined by the two-point correlation
functions
\begin{equation}\nn
\left\langle{\cal O}^{\dagger}_i(x_1){\cal O}_j(x_2)\right\rangle=
\frac{C_{12}\delta_{ij}}{|x_1-x_2|^{2\Delta_i}}
\end{equation}
and $C_{ijk}$ are the structure constants in the operator product
expansion
\begin{equation}\nn
\left\langle{\cal O}_i(x_1){\cal O}_j(x_2){\cal
O}_k(x_3)\right\rangle=
\frac{C_{ijk}}{|x_1-x_2|^{\Delta_1+\Delta_2-\Delta_3}
|x_1-x_3|^{\Delta_1+\Delta_3-\Delta_2}|x_2-x_3|^{\Delta_2+\Delta_3-\Delta_1}}.
\end{equation}
Therefore, the determination of the initial conformal data for a
given conformal field theory is the most important step in the
conformal bootstrap approach.

The three-point functions of two "heavy" operators and a "light"
operator can be approximated by a supergravity vertex operator
evaluated at the "heavy" classical string configuration
\cite{rt10,Hernandez2}: \bea \nn \langle
V_{H}(x_1)V_{H}(x_2)V_{L}(x_3)\rangle=V_L(x_3)_{\rm classical}.
\eea For $\vert x_1\vert=\vert x_2\vert=1$, $x_3=0$, the
correlation function reduces to \bea \nn \langle
V_{H}(x_1)V_{H}(x_2)V_{L}(0)\rangle=\frac{C_{123}}{\vert
x_1-x_2\vert^{2\Delta_{H}}}. \eea Then, the normalized structure
constants \bea \nn \mathcal{C}=\frac{C_{123}}{C_{12}} \eea can be
found from \bea \label{nsc} \mathcal{C}=c_{\Delta}V_L(0)_{\rm
classical}, \eea were $c_{\Delta}$ is the normalized constant of
the corresponding "light" vertex operator.

By now, investigations on the {\it finite-size} effects in the
three-point correlators have been performed in
\cite{AB1105,Lee:2011,AB11062,PLB1107,PLB1108,B1212}. This was
done for the cases when the "heavy" string states are {\it
finite-size} giant magnons, with one or two angular
momenta\footnote{See also \cite{BBP1211}, where the finite-size
correction to a three-point correlation function is found, when
the "heavy" state is not giant magnon one.}, and for three
different "light" states:
\begin{enumerate}
\item{Primary scalar operators: $V_L=V^{pr}_j$} \item{Dilaton
operator: $V_L=V^d_j$} \item{Singlet scalar operators on higher
string levels: $V_L= V^q$}
\end{enumerate}

According to \cite{rt10}, the corresponding (unintegrated)
vertices are given by

\bea \label{prv} V^{pr}_j&=&\left(Y_4+Y_5\right)^{-\Delta_{pr}}
\left(X_1+iX_2\right)^j
\\ \nn
&&\left[z^{-2}\left(\p x_{m}\bar{\p}x^{m}-\p z\bar{\p}z\right) -\p
X_{k}\bar{\p}X_{k}\right],\eea where the scaling dimension is
$\Delta_{pr}=j$. The corresponding operator in the dual gauge
theory is $Tr\left( Z^j\right)$ \footnote{$Z$ is a complex scalar
.}.
\bea \label{dv} V^d_j&=&\left(Y_4+Y_5\right)^{-\Delta_d}
\left(X_1+iX_2\right)^j
\\ \nn
&&\left[z^{-2}\left(\p x_{m}\bar{\p}x^{m}+\p z\bar{\p}z\right) +\p
X_{k}\bar{\p}X_{k}\right], \eea where now the scaling dimension
$\Delta_d=4+j$ to the leading order in the large $\sqrt{\lambda}$
expansion. The corresponding operator in the dual gauge theory is
proportional to $Tr\left(F_{\mu\nu}^2 \ Z^j+\ldots\right)$, or for
$j=0$, just to the SYM Lagrangian. \bea\label{Vq}   V^q=
(Y_4+Y_5)^{- \Delta_q} (\p X_k \bar{\p} X_k)^q .\eea This operator
corresponds to a scalar {\it string} state at level $n=q-1$, and
to the leading order in $\frac{1}{\sqrt{\lambda}}$ expansion
\bea\label{mc}
\Delta_q=2\left(\sqrt{(q-1)\sqrt{\lambda}+1-\frac{1}{2}q(q-1)}+1\right).\eea
The value $n=1 (q=2)$ corresponds to a massive string state on the
first exited level and the corresponding operator in the dual
gauge theory is an operator contained within the Konishi
multiplet. Higher values of $n$ label higher string levels.

In (\ref{prv}), (\ref{dv}), (\ref{Vq}) we denoted with $Y$, $X$
the coordinates in $AdS$ and sphere parts of the $AdS_5\times
S^5_{\gamma}$ background. \bea\nn &&Y_1+iY_2=\sinh\rho\ \sin\eta\
e^{i\varphi_1},\\ \nn &&Y_3+iY_4=\sinh\rho\ \cos\eta\
e^{i\varphi_2},\\ \nn &&Y_5+iY_0=\cosh\rho\ e^{it}. \eea The
coordinates $Y$ are related to the Poincare coordinates by \bea
\nn &&Y_m=\frac{x_m}{z},\\ \nn
&&Y_4=\frac{1}{2z}\left(x^mx_m+z^2-1\right), \\ \nn
&&Y_5=\frac{1}{2z}\left(x^mx_m+z^2+1\right), \eea where $x^m
x_m=-x_0^2+x_ix_i$, with $m=0,1,2,3$ and $i=1,2,3$.

The semiclassical results found in \cite{PLB1107,PLB1108} for the
normalized structure constants (\ref{nsc}), in the case of
finite-size giant magnons on the $\gamma$-deformed $AdS_5\times
S^5_{\gamma}$ and the above three vertices, are given by
\bea\label{cprjg} &&\mathcal{C}_{j\tilde{\gamma}}^{pr}=
\pi^{3/2}c_{j}^{pr}
\frac{\Gamma\left(\frac{j}{2}\right)}{\Gamma\left(\frac{1+j}{2}\right)}
\frac{(1-v^2)\chi_p^{j/2}}{\sqrt{(1-u^2)\left(\chi_p-\chi_n\right)}}
\\ \nn &&
\left\{\left[\sqrt{W}\frac{j-1}{j+1} +\frac{1}{\sqrt{W}(1-v^2)}
\left(2-(1+v^2)W-2\tilde{\gamma}K\right)\right]\right.
\\ \nn && \left.\times
F_1\left(1/2,1/2,-j/2;1;1-\epsilon,1-\frac{\chi_m}{\chi_p}\right)\right.
\\ \nn
&&- \left.\frac{2}{\sqrt{W}(1-v^2)}\left[1-\tilde{\gamma}K
-u\left(u-\tilde{\gamma}u
K+\tilde{\gamma}vW\right)\right]\chi_p\right.
\\ \nn && \left. \times
F_1\left(1/2,1/2,-1-j/2;1;1-\epsilon,1-\frac{\chi_m}{\chi_p}\right)\right\},\eea

\bea\label{cdjg} &&\mathcal{C}_{j\tilde{\gamma}}^{d}=
2\pi^{3/2}c_{4+j}^{d}
\frac{\Gamma\left(\frac{4+j}{2}\right)}{\Gamma\left(\frac{5+j}{2}\right)}
\frac{\chi_p^{j/2}}{\sqrt{(1-u^2)W\left(\chi_p-\chi_n\right)}}
\\ \nn &&
\left\{\left[1-\tilde{\gamma}K-u\left(u+\tilde{\gamma}(vW-uK)\right)\right]\chi_p
F_1\left(1/2,1/2,-1-j/2;1;1-\epsilon,1-\frac{\chi_m}{\chi_p}\right)\right.
\\ \nn
&&-\left.\left(1-W-\tilde{\gamma}K\right)
F_1\left(1/2,1/2,-j/2;1;1-\epsilon,1-\frac{\chi_m}{\chi_p}\right)\right\},\eea

\bea\label{cqg} &&\mathcal{C}_{\tilde{\gamma}}^{q}=
c_{\Delta_q}\pi^{3/2}
\frac{\Gamma\left(\frac{\Delta_q}{2}\right)}{\Gamma\left(\frac{\Delta_q+1}{2}\right)}
\frac{(-2A)^q}{(1-v^2)^{q-1}\sqrt{(1-u^2)W(\chi_p-\chi_n)}}
\\ \nn &&\sum_{k=0}^{q}\frac{q!}{k!(q-k)!}\left(-\frac{B}{A}\right)^{k}\chi_p^{k}\
F_1\left(\frac{1}{2},\frac{1}{2},-k;1;1-\epsilon,1-\frac{\chi_m}{\chi_p}\right),\eea
where \bea\label{deg} &&A=1-\frac{1}{2}(1+v^2)W-\tilde{\gamma}K
,\h B=1-\tilde{\gamma}K -u\left[u-\tilde{\gamma}(K u-v
W)\right],\eea and $F_1(a,b_1,b_2;c;z_1,z_2)$ is one of the
hypergeometric functions of two variables ($Appell F_1$).

\setcounter{equation}{0}
\section{Leading order finite-size corrections}
From now on, we restrict ourselves to the case $\mathcal{J}_2=0$,
$\mathcal{J}_{1}$ large but finite, i.e. $J_1\gg \sqrt{\lambda}$.
This means that the problem reduces to consider the limit
$\epsilon\to 0$, since $\epsilon= 0$ corresponds to the
infinite-size case, i.e. $\mathcal{J}_{1}=\infty$ (see
(\ref{eps0})). To this end, we introduce the expansions
\bea\nn
&&\chi_p=\chi_{p0}+\left(\chi_{p1}+\chi_{p2}\log(\epsilon)\right)\epsilon,
\\ \nn &&\chi_m=\chi_{m0}+\left(\chi_{m1}+\chi_{m2}\log(\epsilon)\right)\epsilon,
\\ \nn &&\chi_n=\chi_{n0}+\left(\chi_{n1}+\chi_{n2}\log(\epsilon)\right)\epsilon,
\\
\label{Dpars} &&v=v_0+\left(v_1+v_2\log(\epsilon)\right)\epsilon, \\
\nn &&u=u_0+\left(u_1+u_2\log(\epsilon)\right)\epsilon,
\\ \nn &&W=W_0+\left(W_1+W_2\log(\epsilon)\right)\epsilon,
\\ \nn &&K=K_0+\left(K_1+K_2\log(\epsilon)\right)\epsilon .\eea
To be able to reproduce the dispersion relation for the
infinite-size giant magnons, we set \bea\label{is}
\chi_{m0}=\chi_{n0}=K_0=0,\h W_0=1.\eea Also, it can be proved
that if we keep the coefficients $\chi_{m2}$, $\chi_{n2}$, $W_2$
and $K_2$ nonzero, the known leading correction to the giant
magnon energy-charge relation in (\ref{0}) will be modified by a
term proportional to $\mathcal{J}_1^2$. That is why we choose
\bea\label{k2} \chi_{m2}= \chi_{n2}=W_2=K_2=0.\eea In addition,
since we are considering for simplicity giant magnons with one
angular momentum ($\mathcal{J}_2=0$), we also set \bea\label{u0}
u_0=0,\eea because the leading term in the $\epsilon$-expansion of
$\mathcal{J}_2$ is proportional to $u_0$. Thus, (\ref{Dpars})
simplifies to \bea\nn
&&\chi_p=\chi_{p0}+\left(\chi_{p1}+\chi_{p2}\log(\epsilon)\right)\epsilon,
\\ \nn &&\chi_m=\chi_{m1}\epsilon,
\\ \nn &&\chi_n=\chi_{n1}\epsilon,
\\
\label{Dpars0} &&v=v_0+\left(v_1+v_2\log(\epsilon)\right)\epsilon, \\
\nn &&u=\left(u_1+u_2\log(\epsilon)\right)\epsilon,
\\ \nn &&W=1+W_1\epsilon,
\\ \nn &&K=K_1\epsilon .\eea

By replacing (\ref{Dpars0}) in (\ref{3eqsd}) and (\ref{de}), one
finds
\bea\label{chi} &&\chi_{p0}=1-v_0^2, \\
\nn &&\chi_{p1}= \frac{v_0}{1-v_0^2}
\Big[v_0\sqrt{(1-v_0^2)^4-4K_1^2(1-v_0^2)}-2(1-v_0^2)v_1 \Big ],\\
\nn &&\chi_{p2}=
-2v_0v_2 ,\\
\nn &&\chi_{m1}=
\frac{(1-v_0^2)^2+\sqrt{(1-v_0^2)^4-4K_1^2(1-v_0^2)}}
{2(1-v_0^2)}, \\
\nn &&\chi_{n1}=
-\frac{(1-v_0^2)^2-\sqrt{(1-v_0^2)^4-4K_1^2(1-v_0^2)}}
{2(1-v_0^2)},
\\ \nn &&W_1=-\frac{\sqrt{(1-v_0^2)^4-4K_1^2(1-v_0^2)}}
{1-v_0^2}.\eea

The expressions for the other parameters in (\ref{Dpars0}) and
(\ref{chi}) can be derived in the following way.

First, we impose the conditions $\mathcal{J}_2=0$ and $p_1$ to be
independent of $\epsilon$. This leads to four equations with
solution \bea\label{uv}
&&v_1=\frac{v_0\sqrt{(1-v_0^2)^4-4K_1^2(1-v_0^2)} \left(1-\log
16\right)}{4(1-v_0^2)}, \\ \nn &&v_2=
\frac{v_0\sqrt{(1-v_0^2)^4-4K_1^2(1-v_0^2)}}{4(1-v_0^2)}, \\ \nn
&&u_1=\frac{K_1v_0\log 4}{1-v_0^2}, \\ \nn &&u_2=
-\frac{K_1v_0}{2(1-v_0^2)},\eea where \bea\label{v0}
v_0=\cos\frac{p_1}{2}.\eea

Second, expanding $\mathcal{J}_1$ and $p_2=2\pi n_2$ $(n_2\in
\mathbb{Z})$\footnote{This follows from the periodicity condition
on $\phi_2$.} to the leading order in $\epsilon$, we obtain
(compare with (\ref{eps0})) \bea\label{ekf}
\epsilon=16\exp\left(-\frac{\mathcal{J}_1}{\sin\frac{p_1}{2}}-2\right),
\h K_1=\frac{1}{2}\sin^3\frac{p_1}{2}\sin\Phi, \h
\Phi=2\pi\left(n_2-\frac{\tilde{\gamma}}{2\pi}\mathcal{J}_1\right).\eea

Now, we are going to use the above results to find the leading
order finite-size effects on the normalized structure constants in
terms of $\mathcal{J}_1\equiv \mathcal{J}$, $p_1\equiv p$ and
$\Phi$.

%%%%%%%%%%%%%5%%%%%%%%%%%%%%%%%%%%%%%%%%%%%%%%%%%%%%%%%%%%%%%%%%%%%%%%%%%%%%%%
\subsection{Giant magnons on $AdS_5 \times S^5_{\gamma}$ and primary scalar
operators}
%%%%%%%%%%%%%5%%%%%%%%%%%%%%%%%%%%%%%%%%%%%%%%%%%%%%%%%%%%%%%%%%%%%%%%%%%%%%%%

As was pointed out in \cite{B1212}, where the undeformed case has
been considered, $j=1$ and $j=2$ are special values. That is why
we will find the corresponding normalized structure constants in
Appendix A. Here, we will deal with $j\ge 3$, when we can use the
following representation of $F_1(a,b_1,b_2;c;z_1,z_2)$ \cite{w}:
\bea\label{F1exp2F1} F_1(a,b_1,b_2;c;z_1,z_2)= \sum_{k=
0}^{\infty} \frac{(a)_{k}(b_2)_{k}}{(c)_{k}}\
{}_2F_1\left(a+k,b_1;c+k;z_1\right)\frac{z_2^k}{k!} .\eea

For all cases we are going to consider, primary, dilaton and
higher string level vertices, only $b_2$ is different, while the
other parameters and arguments of $F_1$ are the same (see
(\ref{cprjg}), (\ref{cdjg}), (\ref{cqg})) : \bea\label{pars}
a=\frac{1}{2},\h b_1=\frac{1}{2},\h c=1,\h z_1=1-\epsilon,\h
z_2=1-\frac{\chi_m}{\chi_p}.\eea

Then, expending
${}_2F_1\left(\frac{1}{2}+k,\frac{1}{2};1+k;1-\epsilon\right)(1-\chi_m/\chi_p)^k$
around $\epsilon=0$, one finds \bea\nn
&&{}_2F_1\left(\frac{1}{2}+k,\frac{1}{2};1+k;1-\epsilon\right)\left(1-\frac{\chi_m}{\chi_p}\right)^k
\approx \frac{\Gamma(1+k)}{\sqrt{\pi}\ \Gamma(\frac{1}{2}+k)}
\left\{\log(4)-H_{k-\frac{1}{2}}\right. \\ \nn &&-\left.
\frac{1}{4\chi_{p0}} \left[2\chi_{p0}+\left(4k
\chi_{m1}-(1+2k)\chi_{p0}\right)\left(\log(4)-H_{k-\frac{1}{2}}\right)\right]\epsilon
-\log(\epsilon)\right.
\\ \label{expF}
&&-\left.\frac{\chi_{p0}+2k(\chi_{p0}-2\chi_{m1})}{4\chi_{p0}}\
\epsilon\log(\epsilon)\right\},\eea where $H_{z}$ is defined as
\cite{w}\bea\nn H_{z}=\psi(z+1)+\gamma.\eea

The replacement of (\ref{expF}) in (\ref{F1exp2F1}), taking into
account (\ref{pars}), gives \bea\label{F1exp}
F_1\left(\frac{1}{2},\frac{1}{2},b_2;1;1-\epsilon,1-\frac{\chi_m}{\chi_p}\right)\approx
C_0 + C_1\ \epsilon + C_2\ \epsilon \log(\epsilon) + C_3
\log(\epsilon),\eea where \bea\label{Ccoeff} &&C_0=
\frac{\Gamma(-b_2)}{\sqrt{\pi}\
\Gamma\left(\frac{1}{2}-b_2\right)} + \frac{\log(16)}{\pi}\
{}_1F_0(b_2,1),\\
\nn &&C_1= \frac{1}{4\pi}\Bigg
\{\frac{1}{\chi_{p0}}\Bigg[-\frac{\sqrt{\pi}\
\Gamma(-1-b_2)}{\Gamma\left(\frac{1}{2}-b_2\right)}\ (\chi_{p0}+2
b_2 \chi_{m1})
\\ \nn &&+8 \log(2)\ b_2 (\chi_{p0}-2 \chi_{m1})\
{}_1F_0(1+b_2,1)\Bigg ]-2(1-\log(4))\ {}_1F_0(b_2,1)\Bigg \},\\
\nn &&C_2=-\frac{1}{4\pi\chi_{p0}} \Big[\chi_{p0}\
{}_1F_0(b_2,1)+2 b_2(\chi_{p0}-2\chi_{m1})\ {}_1F_0(1+b_2,1)\Big],\\
\nn &&C_3= -\frac{1}{\pi}\ {}_1F_0(b_2,1).\eea Here,
${}_1F_0(b,z)$ is one of the hypergeometric functions.

In the normalized structure constants (\ref{cprjg}), there are two
hypergeometric functions
$F_1\left(\frac{1}{2},\frac{1}{2},b_2;1;1-\epsilon,1-\frac{\chi_m}{\chi_p}\right)$
with $b_2=-j/2$ and $b_2=-1-j/2$. By using (\ref{F1exp}),
(\ref{Ccoeff}) in (\ref{cprjg}) and expanding it about
$\epsilon=0$, we can write down the following approximate equality
for $j\ge 3$ \bea\label{cprjgg2}
&&\mathcal{C}_{j\tilde{\gamma}}^{pr}\approx A_0+A_1 \epsilon+A_2
\epsilon\log(\epsilon),\eea where the coefficients are given by
\bea\label{Acoeffs} &&A_0= c_j^{pr}\pi
\frac{\Gamma(\frac{j}{2})^2}{\Gamma(\frac{1+j}{2})\Gamma(\frac{3+j}{2})}
\ j \ \chi_{p0}^{\frac{1}{2}(j-1)} (1-v_0^2-\chi_{p0}),\\ \nn
&&A_1= c_j^{pr}\frac{\pi}{4}
\frac{\Gamma(\frac{j}{2})\Gamma(\frac{j}{2}-1)}{\Gamma(\frac{1+j}{2})\Gamma(\frac{3+j}{2})}
\ \chi_{p0}^{\frac{1}{2}(j-3)}
\Big\{4(W_1+\chi_{m1})\chi_{p0}-2\chi_{p0}^2
\\ \nn
&&-\left[2\chi_{n1}(1-v_0^2-\chi_{p0})+\chi_{p0}(1-v_0(v_0+8v_1+2v_0W_1)\right.
\\ \nn
&&-\left.\chi_{p0}(1-2W_1))-2(1-v_0^2+\chi_{p0})\chi_{p1}\right]j
\\ \nn
&&+\left[\chi_{n1}-4v_0v_1\chi_{p0}+\chi_{m1}(1-v_0^2-\chi_{p0})
-v_0^2(\chi_{n1}+W_1\chi_{p0}-3\chi_{p1})-3\chi_{p1}\right.
\\ \nn
&&+\left.\chi_{p0}\left(-\chi_{n1}+W_1(-1+\chi_{p0})+\chi_{p1}\right)\right]j^2
\\ \nn
&&+(1-v_0^2-\chi_{p0})\chi_{p1}\ j^3
\\ \nn
&&+\tilde{\gamma}\left[4K_1\chi_{p0}+\left(2\chi_{p0}\left(K_1-2(K_1+v_0u_1)\chi_{p0}\right)\right)j
+\left(2\chi_{p0}\left(v_0u_1\chi_{p0}-K_1(1-\chi_{p0})\right)\right)j^2\right]\Big\},\\
\nn &&A_2= -c_j^{pr}\frac{\pi}{2}
\frac{\Gamma(\frac{j}{2})^2}{\Gamma(\frac{1+j}{2})\Gamma(\frac{3+j}{2})}
\ j \
\chi_{p0}^{\frac{1}{2}(j-3)}\left[4v_0v_2\chi_{p0}+(1-v_0^2+\chi_{p0})\chi_{p2}\right.
\\ \nn
&&-\left.(1-v_0^2-\chi_{p0})\chi_{p2}\
j-2\tilde{\gamma}v_0u_2\chi_{p0}^2\right].\eea

Now, our goal is to express (\ref{cprjgg2}) in terms of
$\mathcal{J}$, $p$, and $\Phi$. To this end, we replace
(\ref{chi}) - (\ref{ekf}) in (\ref{Acoeffs}). This leads to the
following final result for the normalized structure constants
$\mathcal{C}_{j\tilde{\gamma}}^{pr}$ for $j\ge 3$ \bea\label{Cprf}
&&\mathcal{C}_{j\tilde{\gamma}}^{pr}\approx c_j^{pr}\frac{\pi}{8}
\frac{\Gamma\left(\frac{j}{2}\right)}{\Gamma\left(\frac{1+j}{2}\right)\Gamma\left(\frac{3+j}{2}\right)}
\sin\left(\frac{p}{2}\right)^{1+j}\left[4(j-1)\Gamma\left(\frac{j}{2}-1\right)\cos(\Phi)\right.
\\ \nn
&&-\left.\tilde{\gamma}\Gamma\left(\frac{j}{2}\right)
\left(4\sin\left(\frac{p}{2}\right)-j\left(1+\cos\left(p\right)\right)\mathcal{J}\right)
\sin(\Phi)\right]\epsilon.\eea Let us point out that (\ref{Cprf})
reduces exactly to the result found for the undeformed case in
\cite{B1212}, when $\tilde{\gamma}=0$, $\Phi=0$. Moreover, it
generalizes it for any $j\ge 3$. The cases $j=1$ and $j=2$ will be
considered separately in Appendix A.

%%%%%%%%%%%%%5%%%%%%%%%%%%%%%%%%%%%%%%%%%%%%%%%%%%%%%%%%%%%%%%%%%%%%%%%%%%%%%%
\subsection{Giant magnons on $AdS_5\times S^5_{\gamma}$ and dilaton operator}
%%%%%%%%%%%%%5%%%%%%%%%%%%%%%%%%%%%%%%%%%%%%%%%%%%%%%%%%%%%%%%%%%%%%%%%%%%%%%%

Since the hypergeometric functions in (\ref{cdjg}) are the same as
in (\ref{cprjg}) (only the coefficients in front of them are
different), we can use (\ref{F1exp}), (\ref{Ccoeff}) for the case
under consideration, with $b_2=-j/2$ and $b_2=-1-j/2$. Thus,
expanding (\ref{cdjg}) to the leading order in $\epsilon$, one
finds ($j\ge 1$) \bea\nn
&&\mathcal{C}_{j\tilde{\gamma}}^{d}\approx c_{4+j}^{d}
\frac{\sqrt{\pi}}{2}
\frac{\Gamma\left(\frac{4+j}{2}\right)}{\Gamma\left(\frac{5+j}{2}\right)}
\chi_{p0}^{\frac{1}{2}(j-1)}\Big\{\epsilon\left[4W_1+(2+j)(2\chi_{m1}-\chi_{p0})
+4\tilde{\gamma}K_1\right]
\\ \nn
&& \times\log\frac{16}{\epsilon}\
{}_1F_0\left(-\frac{j}{2},1\right)
\\ \label{d1}
&&+\sqrt{\pi}\frac{\Gamma\left(\frac{j}{2}\right)}{\Gamma\left(\frac{3+j}{2}\right)}
\Big[2 j
\chi_{p0}+\left(2\chi_{m1}-\chi_{p0}+W_1\left(2+j(2-\chi_{p0})\right)\right.
\\ \nn &&+\left.j(\chi_{m1}+\chi_{n1}+(1+j)\chi_{p1})
+2\tilde{\gamma}(K_1+j(K_1-(K_1+v_0 u_1) \chi_{p0}))\right)\
\epsilon
\\ \nn
&&+(j(1+j)\chi_{p2}-2\tilde{\gamma}jv_0u_2\chi_{p0})\
\epsilon\log\epsilon\Big]\Big\}.\eea

Taking into account (\ref{chi}) - (\ref{ekf}) in (\ref{d1}) one
finally derives \bea\nn &&\mathcal{C}_{j\tilde{\gamma}}^{d}\approx
c_{4+j}^{d}\pi\frac{\Gamma\left(\frac{j}{2}\right)\Gamma\left(\frac{4+j}{2}\right)}
{\Gamma\left(\frac{3+j}{2}\right)\Gamma\left(\frac{5+j}{2}\right)}
\sin^{1+j}(p/2) \Big\{j-\frac{1}{8}\Big[\left(4-j(1+3j)(1+\cos
p)\right.
\\ \nn &&-\left. j(1+j)(1+\cos p)\csc(p/2)\right)\mathcal{J}\cos\Phi
\\ \nn &&-\tilde{\gamma}\left(4\sin(p/2)- j(1+\cos p)\mathcal{J}\right)
\sin\Phi\Big]\epsilon\Big\}.\eea

%%%%%%%%%%%%%5%%%%%%%%%%%%%%%%%%%%%%%%%%%%%%%%%%%%%%%%%%%%%%%%%%%%%%%%%%%%%%%%
\subsection{Giant magnons on $AdS_5\times S^5_{\gamma}$
and singlet scalar operators \\ on higher string levels}
%%%%%%%%%%%%%5%%%%%%%%%%%%%%%%%%%%%%%%%%%%%%%%%%%%%%%%%%%%%%%%%%%%%%%%%%%%%%%%

For this case, we were not able to obtain a general formula for
the leading finite-size corrections to the three-point correlation
functions in terms of $\mathcal{J}$, $p$, and $\Phi$, for any
$q\ge 1$. That is why, we are going to present here the results
for $q=1,...,5$ (string levels $n=0,1,2,3,4$).

Let us first point out that the hypergeometric functions
$F_1\left(\frac{1}{2},\frac{1}{2},-k;1;1-\epsilon,1-\frac{\chi_m}{\chi_p}\right)$
entering (\ref{cqg}) can be expressed in terms of the complete
elliptic integrals $\mathbf{K}(1-\epsilon)$,
$\mathbf{E}(1-\epsilon)$ of first and second kind. For example,
\bea\nn
&&F_1\left(\frac{1}{2},\frac{1}{2},0;1;1-\epsilon,1-\frac{\chi_m}{\chi_p}\right)=
\frac{2}{\pi} \mathbf{K}(1-\epsilon) ,\\ \nn
&&F_1\left(\frac{1}{2},\frac{1}{2},-1;1;1-\epsilon,1-\frac{\chi_m}{\chi_p}\right)=
\frac{2}{\pi}\frac{(\chi_{m}-\epsilon\chi_{p})\mathbf{K}(1-\epsilon)
-(\chi_{m}-\chi_{p})\mathbf{E}(1-\epsilon)}{(1-\epsilon)\chi_p}
\\ \nn
&&F_1\left(\frac{1}{2},\frac{1}{2},-2;1;1-\epsilon,1-\frac{\chi_m}{\chi_p}\right)=
\frac{1}{3\pi(1-\epsilon)^2\chi_p^2}
\Big[2\left((3-\epsilon)\chi_m^2-4\ \epsilon\chi_m\chi_p\right.
\\ \nn
&&-\left.(1-3\ \epsilon)\ \epsilon\chi_p^2
\right)\mathbf{K}(1-\epsilon)-4(\chi_{m}-\chi_{p})\left((2-\epsilon)\chi_m
+ (1-2\ \epsilon)\chi_p\right)\mathbf{E}(1-\epsilon)\Big] .\eea
Then, (\ref{cqg}) can be written in terms of hypergeometric
functions of the type ${}_pF_q$ with argument $1-\epsilon$.
However, this is just much more complicated representation of the
semiclassically exact result.

Here, we are interested in the case of small $\epsilon$ (or,
equivalently, large $\mathcal{J}$) limit. So, we will expand
everything in $\epsilon$. Since the computations are similar to
the previously considered cases, we will write down the final
results only. They are given by the following approximate
equalities: \bea\nn &&\mathcal{C}_{\tilde{\gamma}}^{1}\approx
c_{\Delta_1}\frac{\sqrt{\pi}}{8}\frac{\Gamma\left(\frac{\Delta_1}{2}\right)}
{\Gamma\left(\frac{1+\Delta_1}{2}\right)}\sin(p/2)
\Big\{16-8\mathcal{J}\csc(p/2)+\left[4-\left(2\left(1-\cos p
+\mathcal{J}^2\cot^2(p/2)\right)\right.\right.
\\ \nn
&&+\left.\left.\mathcal{J}(5-\cos
p)\csc(p/2)\right)\cos\Phi+8\tilde{\gamma}\mathcal{J}\sin^2(p/2)
\sin\Phi\right]\epsilon\Big\},\eea

\bea\nn &&\mathcal{C}_{\tilde{\gamma}}^{2}\approx
-c_{\Delta_2}\frac{\sqrt{\pi}}{24}\frac{\Gamma\left(\frac{\Delta_2}{2}\right)}
{\Gamma\left(\frac{1+\Delta_2}{2}\right)}
\Big\{8(2\sin(p/2)-3\mathcal{J})+\Big[12\sin(p/2) \\ \nn &&+
\left(2(27+5\cos p)\sin(p/2)-\mathcal{J}(31+13\cos
p+3\mathcal{J}(1+\cos p)\csc (p/2))\right)\cos\Phi
\\ \nn &&-8\tilde{\gamma}\sin(p/2)(8\sin(p/2)-\mathcal{J}(7+\cos p))\sin\Phi\Big]\epsilon\Big\},\eea

\bea\nn &&\mathcal{C}_{\tilde{\gamma}}^{3}\approx c_{\Delta_3}
\frac{\sqrt{\pi}}{120}\frac{\Gamma\left(\frac{\Delta_3}{2}\right)}
{\Gamma\left(\frac{1+\Delta_3}{2}\right)}
\Big\{8(38\sin(p/2)-15\mathcal{J})+\Big[60\sin(p/2) \\ \nn &&+
\left(18(13+19\cos p)\sin(p/2)-\mathcal{J}(187+97\cos
p+15\mathcal{J}(1+\cos p)\csc (p/2))\right)\cos\Phi
\\ \nn &&-12\tilde{\gamma}\sin(p/2)(48\sin(p/2)-\mathcal{J}(23-7\cos p))\sin\Phi\Big]\epsilon\Big\},\eea

\bea\nn &&\mathcal{C}_{\tilde{\gamma}}^{4}\approx -c_{\Delta_4}
\frac{\sqrt{\pi}}{840}\frac{\Gamma\left(\frac{\Delta_4}{2}\right)}
{\Gamma\left(\frac{1+\Delta_4}{2}\right)} \Big\{1264\sin(p/2)-
840\mathcal{J}+\Big[\sin(p/2)\left(420 \right.
\\ \nn
&&+\left.\left(4730+2054\cos p-\mathcal{J}(1837+1207\cos
p)\csc(p/2)-210\mathcal{J}^2\cot^2(p/2)\right)\cos\Phi \right.
\\ \nn &&-16\left.\tilde{\gamma} \left(424\sin(p/2)-3\mathcal{J}
(79+9\cos p)\right) \sin\Phi\right)\Big]\epsilon\Big\},\eea

\bea\nn &&\mathcal{C}_{\tilde{\gamma}}^{5}\approx c_{\Delta_5}
\frac{\sqrt{\pi}}{2520}\frac{\Gamma\left(\frac{\Delta_5}{2}\right)}
{\Gamma\left(\frac{1+\Delta_5}{2}\right)}
\Big\{8(902\sin(p/2)-315\mathcal{J})+\Big[1260\sin(p/2)
\\ \nn
&&+\left(2(6093+7667\cos
p)\sin(p/2)-\mathcal{J}\left(6343+4453\cos p\right.\right.
\\ \nn
&&+\left.\left.315\mathcal{J}(1+\cos
p)\csc(p/2)\right)\right)\cos\Phi
\\ \nn
&&-20\tilde{\gamma}\sin(p/2)(1376\sin(p/2)-\mathcal{J}(523-107\cos
p))\sin\Phi\Big]\epsilon\Big\}.\eea

%\newpage
\setcounter{equation}{0}
\section{Concluding Remarks}
In this article, we have derived the leading finite-size effects
on the normalized structure constants in some semiclassical
three-point correlation functions in $AdS_5\times S^5_{\gamma}$,
dual to $\mathcal{N} = 1$ SYM theory in four dimensions, arising
as an exactly marginal deformation of $\mathcal{N} = 4$ SYM,
expressed in terms of the conserved string angular momentum
$\mathcal{J}$, and the worldsheet momentum $p$, identified with
the momentum $p$ of the magnon excitations in the dual spin-chain.
More precisely, we found the leading finite-size effects on the
structure constants in three-point correlators of two "heavy"
giant magnon's string states and the following three "light"
states:
\begin{enumerate}
\item{Primary scalar operators}; \item{Dilaton operator with
nonzero-momentum ($j\ge 1$)}; \item{Singlet scalar operators on
higher string levels}.
\end{enumerate}

It would be interesting to investigate other cases for which the
finite-size corrections to the giant magnon's dispersion relations
are known, like $AdS_4\times CP^3$, $AdS_4\times CP^3_{\gamma}$,
$AdS_5\times T^{1,1}$, or $AdS_5\times T^{1,1}_{\gamma}$.

%\section*{Acknowledgements}
%This work was supported in part by DO 02-257 grant.

\def\theequation{A.\arabic{equation}}
\setcounter{equation}{0}
\begin{appendix}

\section{Giant magnons on $AdS_5 \times S^5_{\gamma}$
\\ and primary scalar operators with $j=1$ and $j=2$}

Let us start with the case $j=1$. Expanding the coefficients in
$\mathcal{C}_{1\tilde{\gamma}}^{pr}$ according to (\ref{Dpars0}),
one can rewrite it in the following form \bea\nn
&&\mathcal{C}_{1\tilde{\gamma}}^{pr}\approx
c_1^{pr}\frac{\pi^2}{2} \Bigg\{\frac{1}{\chi_{p0}}
F_1\left(1/2,1/2,-1/2;1;1-\epsilon,1-\frac{\chi_{m1}}{\chi_{p0}}\
\epsilon\right) \Big[\left(1-v_0^2\right)\chi_{n1}\ \epsilon
\\ \nn
&&+\left(2-(4v_0v_1+3W_1+4\tilde{\gamma}K_1)\ \epsilon
-v_0^2(2+W_1\ \epsilon)\right){\chi_{p0}}-4v_0v_2{\chi_{p0}}\
\epsilon\log(\epsilon)\Big]
\\ \label{pr1} &&- F_1\left(1/2,1/2,-3/2;1;1-\epsilon,1-\frac{\chi_{m1}}{\chi_{p0}}\
\epsilon\right)
\\ \nn &&\times
\Big[4\chi_{p0}+2\left(\chi_{n1}-(W_1+2\tilde{\gamma}(K_1+v_0u_1))
\chi_{p0}+2\chi_{p1}\right)\ \epsilon
\\ \nn
&&+4(\chi_{p2}-\tilde{\gamma}v_0u_2\chi_{p0})\ \epsilon
\log(\epsilon)\Big]\Bigg\}.\eea In order to represent
$\mathcal{C}_{1\tilde{\gamma}}^{pr}$ as a function of
$\mathcal{J}$, $p$ and $\Phi$, one have to use (\ref{chi}) -
(\ref{ekf}) in (\ref{pr1}). This leads to \bea\nn
&&\mathcal{C}_{1\tilde{\gamma}}^{pr}\approx -c_1^{pr}
\frac{\pi^2}{4} \sin^2(p/2) \Bigg\{8
F_1\left(1/2,1/2,-3/2;1;1-\epsilon,1-\frac{1}{2}\left(1+\cos\Phi\right)\
\epsilon\right)
\\ \nn &&-4 F_1\left(1/2,1/2,-1/2;1;1-\epsilon,1-\frac{1}{2}\left(1+\cos\Phi\right)\
\epsilon\right)
\\ \label{pr11} &&+\Bigg[F_1\left(1/2,1/2,-1/2;1;1-\epsilon,1-\frac{1}{2}\left(1+\cos\Phi\right)\
\epsilon\right)
\\ \nn &&\times\left(1-\cos\Phi
\left(9+2 \cos p + \mathcal{J}(1+\cos p)\csc(p/2)\right)+4
\tilde{\gamma}\sin(p/2)\sin\Phi\right) \\ \nn
&&-F_1\left(1/2,1/2,-3/2;1;1-\epsilon,1-\frac{1}{2}\left(1+\cos\Phi\right)\
\epsilon\right)
\\ \nn &&\times\left(2-2\cos\Phi\left(5+2\cos p+
\mathcal{J}(1+\cos p)\csc(p/2)\right)\right.
\\ \nn
&&+\left.\tilde{\gamma}\left(\mathcal{J}(1+\cos
p)+4\sin(p/2)\right)\sin\Phi\right)\Bigg]\epsilon\Bigg\}.\eea

For the undeformed case, when $\tilde{\gamma}=0$, $\Phi=0$,
(\ref{pr11}) simplifies to \bea\label{pr1u}
\mathcal{C}_{1}^{pr}\approx -c_1^{pr} \frac{\pi^2}{4} \sin(p/2)
\left[3\sin(p/2)+\sin(3p/2)+\mathcal{J}(1+\cos
p)\right]\epsilon^2.\eea This is in accordance with the result
$\mathcal{C}_{1}^{pr}\approx 0$ found in \cite{B1212}, where only
the leading order in $\epsilon$ was taken into account.

Now, let us consider the case $j=2$, when (\ref{cprjg}) reduces to
\bea\label{pr2} &&\mathcal{C}_{2\tilde{\gamma}}^{pr}= -
\frac{8}{3}c_2^{pr}
\frac{1}{(1-\epsilon)^2\sqrt{(1-u^2)W(\chi_p-\chi_n)}}
\Big\{\Big[3-(1+2v^2)W-3\tilde{\gamma}K\Big](1-\epsilon)
\\ \nn &&\times \Big[\left(\chi_{m}-\chi_{p}\right) \mathbf{E}(1-\epsilon)
-(\chi_{m}-\chi_{p}\ \epsilon)
\mathbf{K}(1-\epsilon)+\left(1-u(u-\tilde{\gamma}(Ku-v W))
-\tilde{\gamma}K\right)
\\ \nn &&\times
(2(\chi_{p}-\chi_{m})((2-\epsilon)\chi_{m}+(1-2\epsilon)\chi_{p})
\mathbf{E}(1-\epsilon)+((3-\epsilon)\chi_{m}^{2}-4\chi_{m}\chi_{p}\
\epsilon
\\ \nn
&&-\chi_{p}^{2}(1-3\epsilon)\
\epsilon)\mathbf{K}(1-\epsilon)\Big]\Big\}.\eea

Expanding (\ref{pr2}) in $\epsilon$, and taking into account
(\ref{chi}) - (\ref{ekf}), one finds \bea\label{pr2f}
&&\mathcal{C}_{2\tilde{\gamma}}^{pr}\approx \frac{2}{3}c_2^{pr}
\sin^2(p/2) \left[2\mathcal{J}\cos\Phi-\tilde{\gamma}
\left(2\sin(p/2)-\mathcal{J}(1+\cos
p)\right)\sin(p/2)\sin\Phi\right]\epsilon.\eea Obviously, the
result for the undeformed case \cite{B1212} is properly reproduced
by the above formula.

\end{appendix}

\end{document}